%% file: tep.tex
\documentclass[pre,twocolumn,showpacs,preprintnumbers,floatfix]{revtex4-1}

\usepackage{etex}
\usepackage{ifpdf}
\usepackage[final,pagebackref]{hyperref}
\usepackage{graphicx}
\usepackage{dcolumn}
\usepackage{url}
\usepackage{xkeyval}
\usepackage{amsfonts}
\usepackage{amsmath}
\usepackage{amssymb}
\usepackage{amsthm}
\usepackage{amscd}
\usepackage{latexsym}
\usepackage{mathtools}
\usepackage{etoolbox}
\usepackage{array}
\usepackage{bm}   
\usepackage{bbm}
\usepackage{verbatim}
\usepackage{color}
\usepackage{xcolor}
\usepackage{setspace}
\usepackage{stmaryrd}
\usepackage[caption=false]{subfig}
\usepackage{tabularx}
\usepackage[capitalize]{cleveref}
\crefname{Prop}{Prop.}{Props.}
\usepackage{nicefrac}
\usepackage[T1]{fontenc}                 
\usepackage{lmodern}                     
\usepackage[scaled=0.72]{beramono}       
\usepackage{microtype}                   
\usepackage{pgf}
\usepackage{tikz}
\usetikzlibrary{arrows}
\usetikzlibrary{automata}
\usetikzlibrary{calc}
\usetikzlibrary{decorations.pathreplacing}
\usetikzlibrary{patterns}
\usetikzlibrary{plotmarks}
\usetikzlibrary{positioning}
\usetikzlibrary{shapes}
\input{vaucanson.tikz}

\usepackage{pgfplots}
\pgfplotsset{compat=newest}
\usepgfplotslibrary{groupplots}

\input{cmechabbrev}

\theoremstyle{plain}
\theoremstyle{plain}
\theoremstyle{plain}
\theoremstyle{plain}\newtheorem{Prop}{Proposition}
\theoremstyle{plain}
\theoremstyle{plain}
\theoremstyle{plain}
\theoremstyle{plain}


\def\clap#1{\hbox to 0pt{\hss#1\hss}}

\def\mathclap{\mathpalette\mathclapinternal}

\def\mathclapinternal#1#2{%
\clap{$\mathsurround=0pt#1{#2}$}}

\newenvironment{itemize*}%
  {\begin{itemize}%
    \setlength{\itemsep}{0pt}%
    \setlength{\parskip}{0pt}%
	\setlength{\topsep}{0pt}%
	\setlength{\partopsep}{0pt}%
	\setlength{\parsep}{0pt}}%
  {\end{itemize}}

\newenvironment{enumerate*}%
  {\begin{enumerate}%
    \setlength{\itemsep}{0pt}%
    \setlength{\parskip}{0pt}%
	\setlength{\topsep}{0pt}%
	\setlength{\partopsep}{0pt}%
	\setlength{\parsep}{0pt}}%
  {\end{enumerate}}

\newcommand{\MS}{\MeasSymbol}
\newcommand{\ms}{\meassymbol}
\newcommand{\CS}{\CausalState}
\newcommand{\cs}{\causalstate}
\newcommand{\AS}{\AlternateState}
\newcommand{\as}{\alternatestate}

\renewcommand{\H}{\operatorname{H}}
\renewcommand{\I}{\operatorname{I}}

\newcommand{\MPast}{\MS_{-m:0}}
\newcommand{\LPresent}{\MS_{0:\ell}}
\renewcommand{\Future}{\MS_{0:}}
\newcommand{\LFuture}{\MS_{\ell:}}
\newcommand{\NFuture}{\MS_{0:n}}
\newcommand{\NLFuture}{\MS_{\ell:n}}

\renewcommand{\past}{\ms_{:0}}

\newcommand{\lpresent}{\ms_{0:\ell}}
\newcommand{\lfuture}{\ms_{\ell:}}

\renewcommand{\hmu}[1]{h_\mu^{#1}}
\renewcommand{\rhomu}[1]{\rho_\mu^{#1}}
\renewcommand{\bmu}[1]{b_\mu^{#1}}
\renewcommand{\rmu}[1]{r_\mu^{#1}}
\renewcommand{\qmu}[1]{q_\mu^{#1}}
\renewcommand{\sigmamu}[1]{\sigma_\mu^{#1}}

\newcommand{\PMI}[1]{\operatorname{PMI(#1)}}

\newcommand{\arxiv}[1]{\href{http://arxiv.org/abs/#1}{\texttt{arXiv}:#1}}
\newcommand{\sfiwp}[1]{Santa Fe Institute Working Paper #1}

\usepackage{listings}
\lstdefinestyle{mypython}{
language=Python,                        
basicstyle=\small\ttfamily,             
keywordstyle=\color{green!50!black},    
commentstyle=\color{gray},              
numbers=left,                           
numberstyle=\tiny,                      
stepnumber=1,                           
numbersep=5pt,                          
backgroundcolor=\color{gray!10},        
frame=none,                             
tabsize=2,                              
captionpos=b,                           
breaklines=true,                        
breakatwhitespace=false,                
showspaces=false,                       
showtabs=false,                         
morekeywords={as},                      
}

\begin{document}

\def\ourTitle{The Elusive Present:\\
Hidden Past and Future Dependency and\\
Why We Build Models}

\author{Pooneh M. Ara}
\email{mohammadiara@ucdavis.edu}
\affiliation{Complexity Sciences Center and Physics Department,
University of California at Davis, One Shields Avenue, Davis, CA 95616}

\author{Ryan G. James}
\email{rgjames@ucdavis.edu}
\affiliation{Complexity Sciences Center and Physics Department,
University of California at Davis, One Shields Avenue, Davis, CA 95616}

\author{James P. Crutchfield}
\email{chaos@cse.ucdavis.edu}
\affiliation{Complexity Sciences Center and Physics Department,
University of California at Davis, One Shields Avenue, Davis, CA 95616}

\def\ourAbstract{%
Modeling a temporal process as if it is Markovian assumes the present encodes
all of the process's history. When this occurs, the present captures all of the
dependency between past and future. We recently showed that if one randomly
samples in the space of structured processes, this is almost never the case.
So, how does the Markov failure come about? That is, how do individual
measurements fail to encode the past? And, how many are needed to capture
dependencies between the past and future? Here, we investigate how much
information can be shared between the past and future, but not be reflected in
the present.  We quantify this elusive information, give explicit calculational
methods, and draw out the consequences. The most important of which is that
when the present hides past-future dependency we must move beyond
sequence-based statistics and build state-based models.
}

\def\ourKeywords{%
 stochastic process, hidden Markov model, causal shielding,
  \texorpdfstring{\eM}{epsilon-machine}, causal states, mutual information.
}

\bibliographystyle{unsrt}

\hypersetup{
  pdfauthor={J. P. Crutchfield},
  pdftitle={\ourTitle},
  pdfsubject={\ourAbstract},
  pdfkeywords={\ourKeywords},
  pdfproducer={},
  pdfcreator={}
}

\begin{abstract}
\ourAbstract

\vspace{0.1in}
\noindent
{\bf Keywords}: \ourKeywords
\end{abstract}

\pacs{
02.50.-r  
89.70.+c  
05.45.Tp  
02.50.Ey  
02.50.Ga  
}

\preprint{\sfiwp{15-07-XXX}}
\preprint{\arxiv{1507.XXXX}}

\title{\ourTitle}
\date{\today}
\maketitle

\setstretch{1.1}

\section{Introduction}

Until the turn of the nineteenth century, temporal processes were almost
exclusively considered to be independently sampled at each time from the same
statistical distribution. These studies were initiated by Jacob Bernoulli in the
1700s \cite{Bern13a} and refined by Simeon Poisson \cite{Pois37a} and Pafnuty
Chebyshev \cite{Cheb67a} in the 1800s, leading to the weak Law of Large Numbers
and the Central Limit Theorem. These powerful results were the first hints at
universal laws in stochastic processes, but they applied only to independent,
identically distributed (IID) processes---unstructured processes with no
temporal correlation, no memory. Moreover, until the turn of the century it was
believed that these laws required independence. It fell to Andrei Andreevich
Markov (1856--1922) to realize that independence is not necessary. To show this
he introduced a new kind of sequence or ``chain'' of dependent random variables,
along with the concepts of transition probabilities, irreducibility, and
stationarity \cite{Mark07a,Mark08a}.

Introducing his ``complex chains'' in 1907, Markov initiated the modern study of
structured, interdependent, and correlated processes. Indeed, in the first and
now-famous application of complex chains, he analyzed the pair distribution
(2-grams) in the $20,000$ vowels and consonants in Pushkin's poem \emph{Eugeny
Onegin} and the $100,000$ letters in Aksakov's novel \emph{The Childhood of
Bagrov, the Grandson} \cite{Mark13a,Mark13b}. Since Markov's time the study of
complex chains has developed into one of the most powerful and widely applied
mathematical theories, far beyond quantitative linguistics to physics, biology,
and finance.

Here, we take an information-theoretic view of Markovian complexity arising from
temporal interdependency between observed symbols that are not themselves the
chain states. Specifically, we consider stationary, ergodic processes generated
by hidden Markov chains (HMCs); introduced in the mid-twentieth
century as a generalization of Markov's chains
necessary to model processes generated by communication channels \cite{Blac57b}.
When are these hidden processes described by finite Markov chains? When are they
not Markovian? What's the informational signature in this case? And, what are
``states'' in the first place? Can we discover them from observations of a
hidden process?

The following is the first in a series that addresses these questions: which
have been answered, which can be answered, and which are open. Here, we
concentrate on how the present---a sequence of $\ell$ consecutive
measurements---statistically shields the past from the future, introducing the
\emph{elusivity} $\sigmamu{\ell}$ as a quantitative measure of shielding. We
show how to calculate it explicitly and then describe and interpret its behavior
(and that of related measures) through examples. As an application we use the
results to reinterpret the persistent mutual information introduced by Ref.
\cite{Ball10a} as a measure of ``emergence'' in complex systems. The sequel
\cite{Ara14b} is analytical, giving closed-form solutions and proving various
properties, including several of those used here.

The next section reviews the minimal necessary background of information theory
\cite{Crut01a}, computational mechanics \cite{Crut12a}, and a recent analysis of
information in the context of the past and future \cite{Jame11a}. We then give
our main new result that expresses the elusivity in terms of a process's causal
states. This leads to simple and efficient expressions, the basis for further
analytical development and empirical estimations. We illustrate the elusivity
for a number of prototype processes and finally compare it to other information
measures. We close with a discussion of the results, drawing conclusions for
future applications.

\section{Information in Complex Processes}
\label{sec:info_theory}

\subsection{Processes}

We are interested in a general stochastic \emph{process} $\Process$: the
distribution of all of a system's behaviors or realizations $\{ \ldots \ms_{-2},
\ms_{-1}, \ms_{0}, \ms_{1}, \ldots \}$ as specified by their joint probabilities
$\Prob(\ldots \MS_{-2}, \MS_{-1}, \MS_{0}, \MS_{1}, \ldots)$. $\MS_t$ is a
random variable that is the outcome of the measurement at the time $t$, taking
values $\ms_t$ from a finite set $\ProcessAlphabet$ of all possible events. We
denote a contiguous chain of random variables as $\LPresent = \MS_0 \MS_1 \cdots
\MS_{\ell-1}$.  Left indices are inclusive; right, exclusive. We suppress
indices that are infinite. We consider only \emph{stationary} processes for
which $\Prob(\MS_{t:t+\ell}) = \Prob(\LPresent)$ for all $t$ and $\ell$.

Our particular emphasis in the following is that a process $\Prob(\Past,
\LPresent, \LFuture)$ is a communication channel that transfers information from
the \emph{past} $\Past = \dots \MS_{-3} \MS_{-2} \MS_{-1}$ to the \emph{future}
$\MS_{\ell:} = \MS_\ell \MS_{\ell+1} \MS_{\ell+2} \dots$ by storing parts of it
in the \emph{present} $\LPresent = \MS_0 \MS_1 \ldots \MS_{\ell-1}$ of length
$\ell$. Of primary concern is whether $\Past \to \LPresent \to \LFuture$ forms a
Markov chain in the sense of Ref.~\cite{Cove06a}:
\begin{align*}
  & \Prob(\MPast, \LPresent, \NLFuture) = \nonumber \\
  & \quad\quad\quad \Prob(\MPast | \LPresent)
   \Prob(\NLFuture | \LPresent) \Prob(\LPresent)
    ~,
\end{align*}
for all $m, n \in \mathbb{Z}^+$.

\subsection{Channel Information}

In analyzing this channel we need to measure the various forms of information
being communicated. The simplest is \emph{Shannon entropy}~\cite{Cove06a}:
\begin{align}
  \H[X] = -\sum_{x \in \mathcal{X}} \Prob(x) \log_2 \Prob(x)
  ~.
\label{eq:entropy}
\end{align}
Three other information-theoretic measures based on the entropy will be
employed throughout. First, the \emph{conditional entropy}, measuring the
amount of information remaining in a variable $X$ (alphabet $\mathcal{X}$)
once the information in a variable $Y$ (alphabet $\mathcal{Y}$) is accounted
for:
\begin{align}
  \H[X|Y] = -\sum_{\substack{x \in \mathcal{X} \\ y \in \mathcal{Y}}} \Prob(x, y) \log_2 \Prob(x | y) ~.
  \label{eq:conditional_entropy}
\end{align}
Second, the deficiency of the conditional entropy relative to the full entropy is known
as the \emph{mutual information}, characterizing the information that is
contained in both $X$ and $Y$:
\begin{align}
  \I[X\!:\!Y] &= \H[X] - \H[X|Y] \nonumber \\
              &= \sum_{\substack{x \in \mathcal{X} \\ y \in \mathcal{Y}}} \Prob(x, y) \log_2 \frac{\Prob(x, y)}{\Prob(x)\Prob(y)} ~.
  \label{eq:mutual_information}
\end{align}
Last, we have the \emph{conditional mutual information}, the mutual information
between two variables once the information in a third ($Z$ with alphabet
$\mathcal{Z}$) has been accounted for:
\begin{align}
  \I[X\!:\!Y|Z] = \sum_{\substack{x \in \mathcal{X} \\ y \in \mathcal{Y} \\ z \in \mathcal{Z}}} \Prob(x, y, z) \log_2 \frac{\Prob(x, y | z)}{\Prob(x | z)\Prob(y | z)}
  ~.
\label{eq:conditional_mutual_information}
\end{align}

Perhaps the most na\"{i}ve way of information-theoretically analyzing a
process, capturing the randomness and dependencies in sequences of random variables, is via the \emph{block entropies}:
\begin{align}
  \H(\ell) = \H[\LPresent]
  ~.
\label{eq:block_entropy}
\end{align}
This quantifies the amount of information in a contiguous block of
observations. Its growth with $\ell$ gives insight into a process's
randomness and structure~\cite{Crut01a, Jame10a}:
\begin{align}
  \H(\ell) \approx \EE + \hmu{} \ell, \quad\quad \ell \gg 1
  ~.
\label{eq:LinearAsymptote}
\end{align}
The asymptotic growth $\hmu{}$, here, is a process's rate of information
generation, or the \emph{Shannon entropy rate}:
\begin{align}
  \hmu{} = \H[\Present|\Past]
  ~.
\label{eq:entropyrate}
\end{align}
And, the amount of future information predictable from the past is the
past-future mutual information or \emph{excess entropy}:
\begin{align}
\label{eq:excess_entropy}
  \EE &= \I[\Past\!:\!\Future] \\
      &= \H[\Past, \Future] - \H[\Past | \Future] ~. \nonumber
\end{align}
The excess entropy naturally arises when considering channels with
a length $\ell = 0$ present,
where it is effectively the only direct information quantity over the variables
$\Past$ and $\Future$. It is well known that if the excess entropy vanishes,
then there is no information temporally communicated by the
channel~\cite{Crut01a}.

Generically, \cref{eq:excess_entropy} is of the form $\infty - \infty$, which
is meaningless. In such situations one refers to finite sequences and then
takes a limit:
\begin{align*}
  \lim_{m, n \to \infty}
 \big( \H[\MPast, \NFuture] - \H[\MPast | \NFuture] \big)
  ~.
\end{align*}
Here, we generally use the informal infinite variables in
equations for clarity and simplicity unless the details of the limit are
important for the analysis at hand. To be concrete, we write $f(\Past)$ to
mean $\displaystyle \lim_{m \to \infty} f(\MPast)$ and $f(\LFuture)$ to mean
$\displaystyle \lim_{n \to \infty} f(\NLFuture)$.

\subsection{Information Atoms}
\label{subsec:atoms}

Our goal here is to analyze a process as a channel as a function of the
present's length $\ell$. The cases of $\ell = 0$ and $\ell = 1$ have already
been addressed: $\ell = 0$ in Ref.~\cite{Crut10a} and $\ell = 1$ in
Ref.~\cite{Jame11a}. Our development closely mirror theirs. We borrow notation,
but must include a superscript to denote the $\ell$-dependence of the
quantities. Our immediate concern is that of monitoring the amount of
dependency remaining between the past and future if the present is known. We
use the mutual information between the past and the future conditioned on the
present to do so---the elusivity that will soon become our focus:
\begin{align}
  \label{eq:sigma_mu}
  \sigmamu{\ell} & = \I[\Past\!:\!\LFuture | \LPresent] \\
                 &= \H[\Past | \LPresent ] + \H[\LFuture | \LPresent]
				 - \H[\Past, \LFuture | \LPresent] \nonumber
  ~.
\end{align}
Note that $\sigmamu{0} = \EE$.

Next, again following Ref.~\cite{Jame11a}, we decompose the length-$\ell$
present. When considering only the past, the information in the present
separates into two components: $\rhomu{\ell} = \I[\Past\!:\!\LPresent]$, the
information that can be anticipated from the past, and $\hmu{\ell} =
\H[\LPresent | \Past]$, the random component that cannot be anticipated.
Naturally, $\H[\LPresent] = \hmu{\ell} + \rhomu{\ell}$. Connecting directly to
Ref.~\cite{Jame11a}, our $\rhomu{1}$ is their $\rhomu{}$ and, likewise,
our $\hmu{1}$ is their $\hmu{}$.

If one also accounts for the future's behavior, then the random, unanticipated
component $\hmu{\ell}$ breaks into two kinds of information: one part
$\bmu{\ell} = \I[\LPresent\!:\!\LFuture | \Past]$ that, while some degree of
randomness, is relevant for predicting the future; and the remaining part
$\rmu{\ell} = \H[\LPresent | \Past, \LFuture]$ is ephemeral, existing only
fleetingly in the present and then dissipating, leaving no trace on future
behavior.

The redundant portion $\rhomu{\ell}$ of $\H[\LPresent]$ itself splits into two
pieces. The first part, $\I[\Past\!:\!\LPresent | \LFuture]$---also $\bmu{\ell}$
when the process is stationary---is shared between the past and the current
observation, but its relevance stops there. The second piece $\qmu{\ell} =
\I[\Past\!:\!\LPresent\!:\!\LFuture]$ is anticipated by the past, is present
currently, and also plays a role in future behavior. Notably, this
informational piece can be negative~\cite{Bell03a, Jame11a}.

Due to a duality between set-theoretic and information-theoretic operators, we
can graphically represent the relationship between these various informations in
a Venn-like display called an \emph{information diagram}~\cite{Yeun08a}; see
\cref{fig:IDiagram}. Similar to a Venn diagram, size indicates Shannon entropy
rather than set cardinality and overlaps are not set intersection, but mutual
information. Each area on the diagram represents one or another of Shannon's
information measures.


\newcommand{\half}{\frac{1}{2}}
\newcommand{\Label}[1]{{\footnotesize \ensuremath{\Symbol{#1}}}}
\newcommand{\edge}[1] {\stackrel{\Symbol{#1}}{\rightarrow}}

\def \r { 1.25cm }

\def \ipast { (-2.5, 0.25) -- +(2,0) arc (90:-90:\r) -- +(-2,0) arc
  (270:90:\r) }
\def \ifuture { (0.5, 0.25) -- +(2,0) arc (90:-90:\r) -- +(-2,0) arc
  (270:90:\r) }
\def \ipresent { (0, 0.25) circle (1.5cm) }
\def \ipresenta { (0, 0) circle (1.511cm) }

\colorlet {past_color}    {red}
\colorlet {pres_color}    {blue}
\colorlet {futu_color}    {black!30!green}

\colorlet {temp_color_1}  {red!50!blue}
\colorlet {temp_color_2}  {red!50!green}
\colorlet {temp_color_3}  {blue!50!green}

\colorlet {hmu_color}     {blue!67!green}
\colorlet {rhomu_color}   {temp_color_1!80!blue}
\colorlet {rmu_color}     {blue}
\colorlet {bmu_1_color}   {temp_color_1}
\colorlet {bmu_2_color}   {temp_color_3}
\colorlet {qmu_color}     {temp_color_1!67!green}
\colorlet {wmu_color}     {temp_color_2!57!blue}
\colorlet {sigmamu_color} {temp_color_2}

\def \opacity { 0.5 }

\pgfdeclarepatternformonly{bigcrosshatch}{\pgfqpoint{-1pt}{-1pt}}{\pgfqpoint{4pt}{4pt}}{\pgfqpoint{5pt}{5pt}}%
  {
  \pgfsetlinewidth{0.4pt}
  \pgfpathmoveto{\pgfqpoint{5.1pt}{0pt}}
  \pgfpathlineto{\pgfqpoint{0pt}{5.1pt}}
  \pgfpathmoveto{\pgfqpoint{0pt}{0pt}}
  \pgfpathlineto{\pgfqpoint{5.1pt}{5.1pt}}
  \pgfusepath{stroke}
  }

\begin{figure}[h!]
  \begin{tikzpicture}[scale=0.95]
    \begin{scope}[even odd rule]
      \clip \ipast;
      \clip \ipast \ipresent;
      \draw [pattern=vertical lines, pattern color=black, opacity=1*\opacity] \ifuture;
    \end{scope}
	  \draw [thick, black] \ipast;
	  \draw [thick, black] \ifuture;
	  \draw [thick, black] \ipresent;
	  \draw (-2.25, 0.6) node {{$\H[\Past]$}};
	  \draw (2.25, 0.6) node {{$\H[\LFuture]$}};
	  \draw (0.0, 2.05) node {{$\H[\LPresent]$}};
	  \draw (0, -1.6) node {{$\sigmamu{\ell}$}};
	  \draw (0, 0.85) node {{$\rmu{\ell}$}};
	  \draw (0, -0.55) node {{$\qmu{\ell}$}};
	  \draw (0.9, -0.2) node {{$\bmu{\ell}$}};
    \draw (-1, -0.2) node {{$\bmu{\ell}$}};
  \end{tikzpicture}
  \caption{
    The process information diagram that places the present in its temporal
    context: the past ($\Past$) and the future ($\LFuture$) partition the
    present ($\LPresent$) into four components with quantities $\rmu{\ell}$,
    $\qmu{\ell}$, and two with $\bmu{\ell}$. Notably, the component
    $\sigmamu{\ell}$, quantifying the hidden dependency shared by the past and
    the future, is \emph{not} part of the present.
  }
\label{fig:IDiagram}
\end{figure}
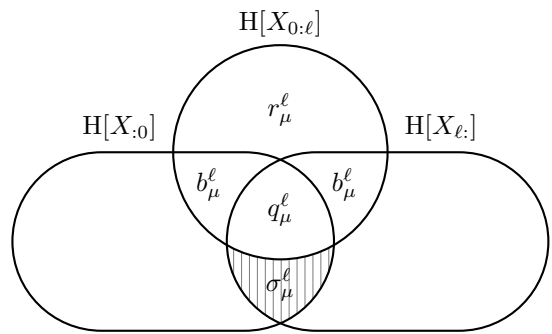

As mentioned above, the past splits $\H[\LPresent]$ yielding two pieces:
$\hmu{\ell}$, the part outside the past, and $\rhomu{\ell}$, the part inside. This
partitioning arises naturally when predicting a process~\cite{Jame11a}. To
emphasize, \cref{subfig:decomp_a} displays this decomposition.  If we include
the future in the diagram, we obtain a more detailed understanding of how
information is transmitted from the past to the future. The past and the future
together divide the present $\H[\LPresent]$ into four parts, as shown in
\cref{subfig:decomp_b}.

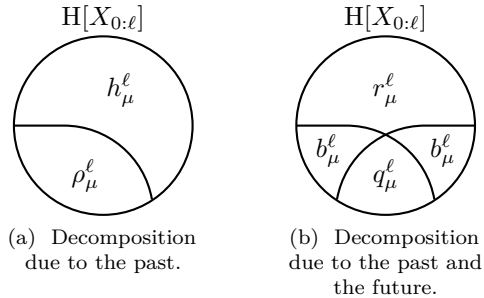
\begin{figure}
  \centering
  \subfloat[
    Decomposition due to the past.
  ]{
    \begin{tikzpicture}[scale=0.95]
      \draw[thick] (0cm,0cm) circle(1.25cm);
      \draw[thick] (-1.25, 0) -- +(0.7,0) arc (90:10:\r);
      \draw (-0.25, -0.7) node {{$\rhomu{\ell}$}};
      \draw (0.25, 0.5) node {{$\hmu{\ell}$}};
      \draw (0, 1.5) node {{$\H[\LPresent]$}};
    \end{tikzpicture}
    \label{subfig:decomp_a}
  } \hspace{1cm}
  \subfloat[
    Decomposition due to the past and the future.
  ]{
    \begin{tikzpicture}[scale=0.95]
      \draw[thick] (0cm,0cm) circle(1.25cm);
      \draw[thick] (-1.25, 0) -- +(0.7,0) arc (90:10:\r);
      \draw[thick]  (1.25, 0) -- +(-0.7,0) arc (90:170:\r);
      \draw (0, 0.5) node {{$\rmu{\ell}$}};
      \draw (0, -0.7) node {{$\qmu{\ell}$}};
      \draw (-0.8, -0.35) node {{$\bmu{\ell}$}};
      \draw (0.8, -0.35) node {{$\bmu{\ell}$}};
      \draw (0, 1.5) node {{$\H[\LPresent]$}};
    \end{tikzpicture}
    \label{subfig:decomp_b}
  }
  \caption{
    Alternative decompositions of the present information $\H[\LPresent]$.
  }
  \label{fig:decompositions}
\end{figure}

The process information diagram makes it rather transparent in which sense
$\rmu{\ell}$ is an amount of \emph{ephemeral information}: its information lies
outside both the past and future and so it exists only in the present moment. It
has no repercussions for the future and is no consequence of the past. It is the
amount of information in the present observation neither communicated to the
future nor from the past. With $\ell = 1$, this has been referred as the
residual entropy rate~\cite{Abda10a}, as it is the amount of uncertainty that
remains in the present even after accounting for every other variable in the
time series. It has also been studied as the erasure
information~\cite{Verdu2008} (there $H^-$), as it is the information
irrecoverably erased in a binary erasure channel.

The \emph{bound information} $\bmu{\ell}$ is the amount of spontaneously
generated information present now, not explained by the past, but that has
consequences for the future. In this sense it hints at being a measure of
structural complexity~\cite{Abda10a,Jame11a}, though we discuss more direct
measures of structure shortly.

Due to stationarity, the mutual information $\I[\LPresent\!:\!\LFuture | \Past]$
between the present $\LPresent$ and the future $\LFuture$ conditioned on the
past $\Past$ is the same as the mutual information $\I[\LPresent\!:\!\Past |
\LFuture]$ between the present $\LPresent$ and the past $\Past$ conditioned on
the future $\LFuture$. Therefore they are both of size $\bmu{\ell}$, as shown in
\cref{fig:IDiagram}. This lends a symmetry to the process information diagram
that need not exist for nonstationary processes.

\subsection{Elusivity}
\label{subsec:elusivity}

Two components remain in the process information diagram---two that have not
been significantly analyzed previously. The first is $\qmu{\ell} =
\I[\Past\!:\!\LPresent\!:\!\LFuture]$---the three-way mutual information (or
\emph{co-information}~\cite{Bell03a}) shared by the past, present, and future.
Notably, unlike Shannon entropies and two-way mutual information, $\qmu{\ell}$
(and co-informations in general) can be negative. The other component
$\sigmamu{\ell} = \I[\Past\!:\!\LFuture | \LPresent]$, the quantity of primary
interest here (shaded in \cref{fig:IDiagram}) is the information shared between
the past and the future that does not exist in the present. Since it measures
dependency hidden from the present, we call it the \emph{elusive information}
or \emph{elusivity} for short.
Generally, it indicates that a process has hidden structures that are not
appropriately captured by finite random-variable blocks. In this case, and as we
discuss at length towards the end, one must build models whose elements,
which we call ``states'' below, represent how a process's internal mechanism is
organized.

A process's internal organization somehow must store all the information from
the past that is relevant for generating the future behavior. Only when the
observed process is Markovian is it sufficient to keep the track of just the
current observable or block of observables. For the general case of
non-Markovian processes, though, information relevant for prediction is spread
arbitrarily far back in the process's history and so cannot be captured by the
present regardless of its duration. This fact is reflected in the existence of
$\sigmamu{\ell}$. When $\sigmamu{\ell} > 0$ for all $\ell$, the description of
the process requires determining its internal organization. This is one reason
to build a model of the mechanism that generates sequences rather than simply
describe a process as a list of sequences.

There are two basic properties that indicate the elusivity's importance. The
first is that $\sigmamu{\ell}$ decreases monotonically as a function of the
present's length $\ell$. That is, dependency cannot increase if we interpolate
more random variables between the past and future.

\begin{Prop}
  $\sigmamu{\ell} \geq \sigmamu{\ell'}$, if $\ell' > \ell$.
\label{prop:decreasing}
\end{Prop}

The second property is that it indicates how poorly the present $\LPresent$
shields the past $\Past$ and future $\LFuture$. When it does, they are
conditionally independent, given the present, and $\sigmamu{\ell}$ vanishes.
Due to this, it can be used to detect a process's \emph{Markov order}
$\MOrder$: the smallest $\MOrder$ for which $\Prob(\MS_0|\Past) =
\Prob(\MS_0|\MS_{-\MOrder:0})$.

\begin{Prop}
  $\sigmamu{\ell} = 0 \iff \ell \geq \MOrder$.
  \label{prop:zero_at_markov_order}
\end{Prop}

Proofs are given in Ref. \cite{Ara14b}.

To calculate $\sigmamu{\ell}$, recall its definition as a conditional
mutual information:
\begin{align*}
  \sigmamu{\ell} &= \I[\Past\!:\!\LFuture | \LPresent] \\
                 &= \sum_{\mathclap{\substack{\past \in \Past \\
                                              \lpresent \in \LPresent \\
                                              \lfuture \in \LFuture}}}
                    \Prob(\ms_{:})
                    \log_2 \frac{\Prob(\past,\lfuture|\lpresent)}
                                {\Prob(\past | \lpresent)\Prob(\lfuture|\lpresent)}
  ~,
\end{align*}
where we used the notational shorthand for the bi-infinite joint distribution
$\Prob(\ms_{:}) = \Prob(\past,\lpresent,\lfuture)$.

Note that for an order-$\MOrder$ Markov process, if $\ell \geq \MOrder$ the past
and the future are independent over range $\MOrder$~\cite{Jame10a} and so
$\Prob(\Past, \LFuture | \LPresent) = \Prob(\Past | \LPresent) \Prob(\LFuture |
\Past)$. With this, it is clear that $\sigmamu{\ell}$ vanishes in such cases.
This current proposition has been discussed in prior literature as well
\cite{Gmei12a}.

Anticipating the needs of our calculations later, we replace conditional
distributions with the joint ones: $\Prob(\past, \lfuture | \lpresent) =
\Prob(\ms_{:}) / \Prob(\lpresent)$ and $\Prob(\past|\lpresent) =
\Prob(\past, \lpresent) / \Prob(\lpresent)$, obtaining:
\begin{align}
  \sigmamu{\ell} = \sum_{\mathclap{\substack{\past \in \Past \\
                                             \lpresent \in \LPresent \\
                                             \lfuture \in \LFuture}}}
                   \Prob(\ms_{:})
                   \log_2 \frac{\Prob(\lpresent) \Prob(\ms_{:})}
                               {\Prob(\past, \lpresent)\Prob(\lpresent, \lfuture)}
  ~.
  \label{eq:sigma_mu_expanded}
\end{align}
Notably, all the terms needed to compute $\sigmamu{\ell}$ are either
$\Prob(\past, \lpresent, \lfuture)$ or marginals thereof. Our next goal,
therefore, is to develop the theoretical infrastructure necessary to
compute that distribution in closed form.

Similar expressions, which we use later on but do not record here, can
be developed for the other information measures $\hmu{\ell}$, $\rmu{\ell}$,
$\bmu{\ell}$, and $\qmu{\ell}$.

\section{Structural Complexity}
\label{sec:structural_complexity}

To analytically calculate the elusive information $\sigmamu{\ell}$ we must go
beyond the information theory of sequences and introduce \emph{computational
mechanics}, the theory of process structure~\cite{Crut12a}. The representation
it uses for a given process is a form of \emph{hidden Markov model}
(HMM)~\cite{Rabi89a}: the \emph{\eM}, which consists of a set $\CausalStateSet$
of \emph{causal states} and a transition dynamic $T$. \EMs\ satisfy three
conditions: irreducibly, unifilarity, and probabilistically distinct
states~\cite{Trav11a}. \emph{Irreducibly} implies that the associated
state-transition graph is strongly connected. \emph{Unifilarity}, perhaps the
most distinguishing feature, means for each state $\cs \in \CausalStateSet$ and
each observed symbol $\ms$ there is at most one outgoing transition from $\CS$
labeled $\ms \in \ProcessAlphabet$. Critically, unifilarity enables one to
directly calculate various process quantities, such as conditional mutual
informations, using properties of the hidden states. Notably, many of these
quantities cannot be directly calculated using the states of general
(nonunifilar) HMMs. Finally, an HMM has \emph{probabilistically distinct
states} when, for every pair of states $\CS$ and $\CS^\prime$, there exists a
word $w$ such that the probability of observing $w$ from each state is
distinct: $\Prob(w|\CS) \neq \Prob(w|\CS^\prime)$. An irreducible, unifilar
model with probabilistically distinct states is minimal in the sense that no
model with fewer states or transitions generates the process. An HMM satisfying
these three properties is an \eM.

\subsection{Constructing the \EM}
\label{subsec:ems}

Given a process, how does one construct it's \eM? First, a process's
\emph{forward causal states}:
\begin{align}
\CausalStateSet^{\forward} & = \MS_{:} / \CausalEquivalence^{\forward}
\label{eq:forward_causal_states}
\end{align}
is the partition defined via the \emph{causal equivalence relation}:
\begin{align}
  \ms_{:t} \CausalEquivalence^{\forward} \ms_{:t}^\prime
    & \equiv \Prob(\MS_{t:} | \MS_{:t}
	  = \ms_{:t}) = \Prob(\MS_{t:} | \MS_{:t} = \ms_{:t}^\prime)
  ~.
\label{eq:forward_causal_equivalence}
\end{align}
That is, each causal state $\cs^{\forward} \in \CausalStateSet^{\forward}$ is
an element of the coarsest partition of a process's pasts such that every
$\ms_{:0} \in \cs^+$ makes the same prediction $\Prob(\MS_{0:} | \cdot)$. In
fact, the causal states are the \emph{minimal sufficient statistic} of the past
to predict the future. We define the \emph{reverse causal states}:
\begin{align}
  \CausalStateSet_t^{\reverse} & = \MS_{t:} / \CausalEquivalence^{\reverse}
  ~.
\label{eq:reverse_causal_states}
\end{align}
by similarly partitioning the process's futures:
\begin{align}
  \ms_{t:} \CausalEquivalence^{\reverse} \ms_{t:}^\prime
    & \equiv \Prob(\MS_{:t} | \MS_{t:} = \ms_{t:})
	= \Prob(\MS_{:t} | \MS_{t:} = \ms_{t:}^\prime)
  ~.
\label{eq:reverse_causal_equivalence}
\end{align}

Second, the causal equivalence relation provides a natural unifilar dynamic
over the states. For each state $\cs$ and next symbol $\ms$, either there is a
successor state $\cs^\prime$ such that the updated past $\ms_{:t+1} = \ms_{:t}
\ms \in \cs^\prime$, for all $\ms_{:t} \in \cs$, or $\ms_{:t+1}$ does not
occur. Due to causal-state equivalence, every past within a state collectively
either can or cannot be followed by a given symbol. Moreover, since the causal
states form a partition of all pasts, there is at most one causal state to
which each past can advance.

For an HMM with states $\as \in \AS$, its \emph{symbol-labeled transition matrix} elements are the probabilities of going from state $\as$ to state $\as^\prime$ and generating the symbol $\ms$:
\begin{align}
  T_{\as\as^\prime}^{(\ms)} \equiv \Prob(\MS_t = \ms, \AS_{t+1} = \as^\prime | \AS_t = \as) ~.
  \label{eq:transition_matrix}
\end{align}
Furthermore, the internal-state dynamics is governed by that stochastic matrix
$\displaystyle T = \sum_{\ms} T^{(\ms)}$. Its unique left eigenvector $\pi$,
associated with eigenvalue 1, gives the asymptotic state probability
$\Prob(\as)$. By extension, the transition matrix giving the probability of a
word $w = \ms_0\ms_1\cdots\ms_{\ell-1}$ of length $\ell$ is the product of
transition matrices of each symbol in $w$:
\begin{align}
  T^{(w)} \equiv \prod_{\ms_i \in w} T^{(\ms_i)} = T^{(\ms_0)} T^{(\ms_1)} \cdots T^{(\ms_{\ell-1})} ~.
  \label{eq:word_transition_matrix}
\end{align}

\def \ipresent { (0, 0.4) circle (1.5cm) }
\def \icausal {  (-1.5, 0.25) -- +(1, 0) arc (90:-90:\r) -- +(-1, 0) arc
  (270:90:\r) }
\def \iicausal {  (0.5, 0.25) -- +(1, 0) arc (90:-90:\r) -- +(-1, 0) arc
  (270:90:\r) }

\begin{figure}
  \centering
  \begin{tikzpicture}[scale=0.75]
    \begin{scope}[even odd rule]
      \clip \ipast;
      \draw [pattern=vertical lines, pattern color=black, opacity=1*\opacity] \ipresent;
    \end{scope}
    \draw [thick, black] \ipast;
    \draw [thick, black] \ifuture;
    \draw [thick, black] \ipresent;
    \draw [ultra thick, black] \icausal;
    \draw [ultra thick, black] \iicausal;
    \draw (-2.25, 0.6) node {{$\H[\Past]$}};
    \draw (0.0, 2.2) node {{$\H[\LPresent]$}};
    \draw (2.25, 0.6) node {{$\H[\LFuture]$}};
    \draw (-1.35, -1.7) node {{$\H[\mathcal{S}^{+}_{0}]$}};
    \draw (1.35, -1.7) node {{$\H[\mathcal{S}^{-}_{\ell}]$}};
  \end{tikzpicture}
\caption{Mutual information $\I[\Past : \LPresent]$ between the past and
  the present (shaded) is equivalent to the mutual information
  $\I[\CS^+_0 : \LPresent]$ between the forward causal state and the present.
  }
\label{fig:miidiagram}
\end{figure}
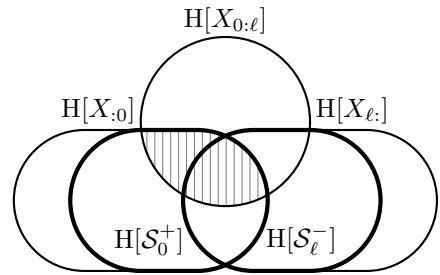

\subsection{Rendering $\sigmamu{\ell}$ Finitely Computable}
\label{subsec:finite}

We can put the forward and reverse causal states to use since they are proxies
for a process's semi-infinite pasts and futures, respectively. See, e.g., Fig.
\ref{fig:miidiagram}. In this way, we transform \cref{eq:sigma_mu} into a form
containing only finite sets of random variables. We calculate directly:
\begin{align}
  \sigmamu{\ell} & = \I[\Past : \LFuture | \LPresent] \nonumber \\
                 & = \I[\Past : (\LPresent, \LFuture)] - \I[\Past : \LPresent] \nonumber \\
                 & = \I[\Past : \Future] - \I[\Past : \LPresent] \nonumber \\
                 & \stackrel{(a)}{=} \I[\CS^+_0 : \CS^-_0] - \I[\CS^+_0 : \LPresent] \nonumber \\
                 & = \I[\CS^+_0 : \CS^-_0] - (\I[\CS^+_0 : \LPresent : \CS^-_0] + \I[\CS^+_0 : \LPresent | \CS^-_0]) \nonumber \\
                 & \stackrel{(b)}{=} \I[\CS^+_0 : \CS^-_0] - \I[\CS^+_0 : \LPresent : \CS^-_0] \nonumber \\
                 & = \I[\CS^+_0 : \CS^-_0 | \LPresent] \nonumber \\
                 & = \I[\CS^+_0 : \CS^-_0 : \CS^-_\ell | \LPresent] + \I[\CS^+_0 : \CS^-_0 | \LPresent, \CS^-_\ell] \nonumber \\
                 & \stackrel{(c)}{=} \I[\CS^+_0 : \CS^-_0 : \CS^-_\ell | \LPresent] \nonumber \\\
                 & = \I[\CS^+_0 : \CS^-_\ell | \LPresent] - \I[\CS^+_0 : \CS^-_\ell | \LPresent, \CS^-_0]  \nonumber \\
                 & \stackrel{(d)}{=} \I[\CS^+_0 : \CS^-_\ell | \LPresent]
				 \nonumber \\
				 & \quad - (\H[\CS^+_0 | \LPresent, \CS^-_0] - \H[\CS^+_0 | \CS^-_0, \LPresent, \CS^-_\ell]) \nonumber \\
                 & = \I[\CS^+_0 : \CS^-_\ell | \LPresent]
  ~.
\label{eq:sigma_mu_states}
\end{align}
Above, $(a)$ is true due to
\crefrange{eq:forward_causal_equivalence}{eq:reverse_causal_states} and
Ref.~\cite{Crut10d}, $(b)$ is true due to
\cref{eq:reverse_causal_equivalence,eq:reverse_causal_states}, $(c)$ is true due
to \cref{eq:reverse_causal_equivalence} and unifilarity, and finally $(d)$ is
true due to both entropy terms being equal to $\H[\CS^+_0 | \CS^-_0]$ by
\cref{eq:reverse_causal_equivalence,eq:reverse_causal_states}. That is,
$\CS^-_0$ informationally subsumes both $\LPresent$ and $\CS^-_\ell$ when it
comes to $\Past$ and, therefore, also when it comes to $\CS^+_0$. All other
equalities are basic information identities found in Ref.~\cite{Cove06a}.

In this way, \cref{eq:sigma_mu_states} says that \cref{eq:sigma_mu_expanded}
becomes, in terms of causal states, a new expression for elusivity:
\begin{align}
  \sigmamu{\ell} = \sum_{\mathclap{\substack{\cs_0^+ \in \CS_0^+ \\
                                             \lpresent \in \LPresent \\
                                             \cs_\ell^- \in \CS_\ell^-}}}
                   \Prob(\cs_0^+, \lpresent, \cs_\ell^-)
                   \log_2 \frac{\Prob(\lpresent) \Prob(\cs_0^+, \lpresent, \cs_\ell^-)}
                               {\Prob(\cs_0^+, \lpresent)\Prob(\lpresent, \cs_\ell^-)}
  ~.
  \label{eq:sigma_mu_states_expanded}
\end{align}
We transformed the key distribution $\Prob(\past, \lpresent, \lfuture)$ over
random variables $\Past$ and $\LFuture$ with cardinality of the continuum to
$\Prob(\cs_0^+, \lpresent, \cs_\ell^-)$ over $\CS^+$ and $\CS^-$ with typically
smaller cardinality. When the causal states are finite or countably infinite,
the benefit is substantial. We will now turn our attention to computing this
joint distribution.

Since the distribution is over both forward and reverse causal states, we must
track both simultaneously. The key tool for this is Ref.~\cite{Crut08b}'s bidirectional machine or \emph{bimachine}. We point the reader there for
details regarding their construction and properties. One feature we need
immediately, though, is that bimachine states $\as_t = (\cs_t^+, \cs_t^-)$ are pairs of forward and reverse causal states.

Generally, given an HMM with states $\as \in \AS$, we can construct the
distribution of interest if we can find a way to build distributions of the
form $\Prob(\as_i, w, \as_j)$: the probability of being in state $\as_i$,
generating the word $w$, and ending in state $\as_j$. The word transition
matrix (Eq.  \eqref{eq:word_transition_matrix}) gives exactly this and allows
us to build the distribution directly:
\begin{align}
  \Prob(\as_i, w, \as_j)=(\pi \circ \textbf{1}_{i}) T^{(w)} \textbf{1}^\intercal_{j} ~,
  \label{eq:distribution}
\end{align}
where $\as_i$ and $\as_j$ are the states of an arbitrary HMM, $a \circ b$ is
the Hadamard (elementwise) product of vectors $a$ and $b$, and $\textbf{1}_{i}$
is the row vector with all its elements zero except for the $i^\textrm{th}$,
which is $1$.

Applying \cref{eq:distribution} to the bimachine, we arrive at the distribution
$\Prob((\CS_0^+, \CS_0^-), \LPresent, (\CS_\ell^+, \CS_\ell^-))$, which can be
marginalized to $\Prob(\CS_0^+, \LPresent, \CS_\ell^-)$, the distribution
needed to compute \cref{eq:sigma_mu_states_expanded}. Figure
\ref{fig:causal_lattice} illustrates this distribution in the setting of the
process's random variable lattice and the forward and reverse causal state
processes.

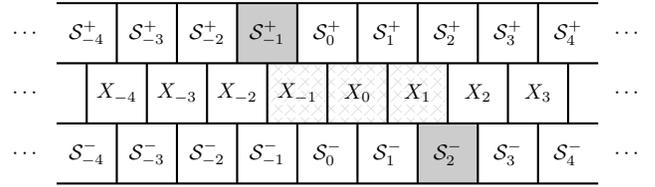
\begin{figure}
  \centering
  \begin{tikzpicture}[scale=0.8, every node/.style={transform shape}]

    \def \Na {9}
    \def \Nb {8}
    \def \diff {4}

    \fill [opacity=0.2] (3, 2) rectangle (4, 3);
    \fill [opacity=0.2] (6, 0) rectangle (7, 1);
    \draw [pattern=bigcrosshatch, opacity=0.33] (3.5, 1) rectangle (6.5, 2);

    \foreach \y in {0,...,3}
      \draw [thick] (0, \y) -- (\Na, \y);
    \foreach \x in {-0.5, \Na+0.5}
      \foreach \y in {0.5,...,2.5}
        \draw (\x, \y) node {$\cdots$};
    \foreach \x in {1,...,\Nb}
      \foreach \y in {0, 2}
        \draw [thick] (\x, \y) -- (\x, \y+1);
    \foreach \x in {0,...,\Nb}
      \draw [thick] (\x+0.5, 1) -- (\x+0.5, 2);
    \foreach \x [evaluate={\n=int(\x-\diff)}] in {0,...,\Nb} {
      \draw (\x+0.5, 2.5) node {$\CausalState^\forward_{\n}$};
      \draw (\x+0.5, 0.5) node {$\CausalState^\reverse_{\n}$};
    }
    \foreach \x [evaluate={\n=int(\x-\diff-1)}] in {1,...,\Nb}
      \draw (\x, 1.5) node {$\MeasSymbol_{\n}$};
  \end{tikzpicture}
\caption{Random variable lattice illustrating the relationship between forward
  causal states $\CS_t^+$, observed symbols $\MS_t$, and reverse causal states
  $\CS_t^-$. The variables in the distribution $\Prob(\CS_{-1}^+, \MS_{-1:2},
  \CS_2^-)$ are highlighted. In particular, elusivity $\sigmamu{3}$ is the
  mutual information between the two shaded cells ($\CS_{-1}^+$ and $\CS_2^-$)
  conditioned on the hatched cells ($\MS_{-1:2} = \MS_{-1} \MS_0 \MS_1$).
  }
  \label{fig:causal_lattice}
\end{figure}

\subsection{Companion Atoms}

Causal-state expressions for $\hmu{\ell}$, $\rmu{\ell}$, $\bmu{\ell}$, and
$\qmu{\ell}$ that we use in the following are:
\begin{align*}
\hmu{\ell} & = \H[\LPresent|\CS^+_0] ~,\\
\rmu{\ell} & = \H[\LPresent|\CS^+_0 : \CS^-_\ell] ~,\\
\bmu{\ell} & = \I[\LPresent:\CS^+_0 | \CS^-_\ell] ~,~\text{and}\\
\qmu{\ell} & = \I[\CS^+_0:\LPresent:\CS^-_\ell]
  ~.
\end{align*}
These are derived in ways paralleling that above for $\sigmamu{\ell}$ and so we
do not give detail. They, too, also depend on the joint distribution above
in Eq. (\ref{eq:distribution}) and its marginals.

\section{Examples}
\label{sec:examples}

Let's consider several example processes, to illustrate calculation methods and
to examine the behavior of $\sigmamu{\ell}$ and companion measures.

\subsection{Golden Mean Process}
\label{subsec:golden_mean}

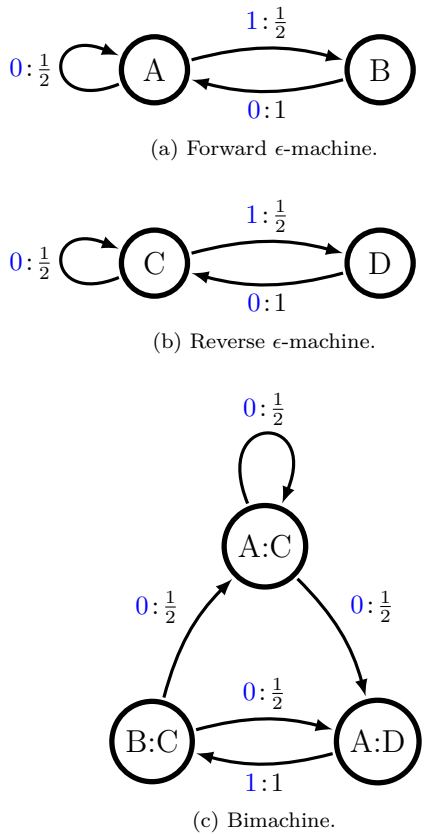
\begin{figure}
  \centering
  \subfloat[Forward \eM.]{
    \centering
    \begin{tikzpicture}[style=vaucanson, bend angle=15, scale=1, every node/.style={transform shape}]
      \node [state] (A)              {A};
      \node [state] (B) [right of=A] {B};

      \path (A) edge [loop left] node {$\Edge{\half}{0}$} (A)
            (A) edge [bend left] node {$\Edge{\half}{1}$} (B)
            (B) edge [bend left] node {$\Edge{1}{0}$}     (A);
      \path (B) edge [loop right, draw=none] node [opacity=0] {$\Edge{0}{0}$} (B);
    \end{tikzpicture}
    \label{subfig:golden_mean_forward}
  } \\
  \subfloat[Reverse \eM.]{
    \centering
    \begin{tikzpicture}[style=vaucanson, bend angle=15, scale=1, every node/.style={transform shape}]
      \node [state] (C)              {C};
      \node [state] (D) [right of=A] {D};

      \path (C) edge [loop left] node {$\Edge{\half}{0}$} (C)
            (C) edge [bend left] node {$\Edge{\half}{1}$} (D)
            (D) edge [bend left] node {$\Edge{1}{0}$}     (C);
      \path (D) edge [loop right, draw=none] node [opacity=0] {$\Edge{0}{0}$} (D);
    \end{tikzpicture}
    \label{subfig:golden_mean_reverse}
  } \\
  \subfloat[Bimachine.]{
    \centering
    \begin{tikzpicture}[style=vaucanson, bend angle=15, scale=1, every node/.style={transform shape}]

      \node [state] (AC) at (0, 0)  {A:C};
      \node [state] (BC) at (240:3) {B:C};
      \node [state] (AD) at (-60:3) {A:D};

      \path (AC) edge [loop above] node {$\Edge{\half}{0}$} (AC)
            (AC) edge [bend left]  node {$\Edge{\half}{0}$} (AD)
            (BC) edge [bend left]  node {$\Edge{\half}{0}$} (AC)
            (BC) edge [bend left]  node {$\Edge{\half}{0}$} (AD)
            (AD) edge [bend left]  node {$\Edge{1}{1}$}     (BC);
    \end{tikzpicture}
    \label{subfig:golden_mean_bidirectional}
  }
  \caption{The several faces of the Golden Mean (GM) Process.}
  \label{fig:golden_mean_process}
\end{figure}

As the first example we analyze the Golden Mean (GM) Process, whose \eMs\ and
bimachine state-transition diagrams are given in
\cref{fig:golden_mean_process}. The GM Process consists of all bi-infinite strings such that no consecutive $1$s occur, with probabilities such that either symbol is equally likely following a $0$. A stochastic generalization of subshifts of finite type~\cite{Lind95a} this process can be described by a Markov chain with order $\MOrder = 1$. Due to
\cref{prop:zero_at_markov_order} we expect $\sigmamu{1} = 0$. To verify this,
we compute each term of \cref{eq:sigma_mu_states_expanded} using the
edges of the bimachine, \cref{subfig:golden_mean_bidirectional}, and
the invariant state distribution $\pi = \big( \nicefrac{1}{3},
\nicefrac{1}{3}, \nicefrac{1}{3} \big)$:
\begin{align*}
  \frac{\Prob(0)\Prob(A, 0, C)}{\Prob(A, 0)\Prob(0, C)} & =
  \frac{\nicefrac{2}{3}\cdot\nicefrac{1}{6}}{\nicefrac{1}{3}\cdot\nicefrac{1}{3}}
  = 1 ~, \\
  \frac{\Prob(0)\Prob(A, 0, D)}{\Prob(A, 0)\Prob(0, D)} & =
  \frac{\nicefrac{2}{3}\cdot\nicefrac{1}{6}}{\nicefrac{1}{3}\cdot\nicefrac{1}{3}}
  = 1 ~, \\
  \frac{\Prob(0)\Prob(B, 0, C)}{\Prob(B, 0)\Prob(0, C)} & =
  \frac{\nicefrac{2}{3}\cdot\nicefrac{1}{6}}{\nicefrac{1}{3}\cdot\nicefrac{1}{3}}
  = 1 ~, \\
  \frac{\Prob(0)\Prob(B, 0, D)}{\Prob(B, 0)\Prob(0, D)} & =
  \frac{\nicefrac{2}{3}\cdot\nicefrac{1}{6}}{\nicefrac{1}{3}\cdot\nicefrac{1}{3}}
  = 1 ~, ~\text{and}\\
  \frac{\Prob(1)\Prob(A, 1, C)}{\Prob(A, 1)\Prob(1, C)} & = \frac{\nicefrac{1}{3}\cdot\nicefrac{1}{3}}{\nicefrac{1}{3}\cdot\nicefrac{1}{3}} = 1
  ~.
\end{align*}
We see that the argument of each $\log_2$ in
\cref{eq:sigma_mu_states_expanded} is $1$, confirming that $\sigmamu{1} = 0$.

\subsection{Information Measures versus Present Length}
\label{subsec:behavior}

We now investigate the behavior of $\sigmamu{\ell}$ and its companions
$\qmu{\ell}$, $\bmu{\ell}$, and $\rmu{\ell}$ for several example processes: the
aforementioned GM, the Even, the Noisy Period Three (NP3), and the
Noisy Random Phase-Slip (NRPS) Processes. The \eMs\ for the latter are shown in
\cref{fig:ems}. Each exhibits different convergence behaviors with $\ell$ for
the differing measures; see the graphs in \cref{fig:growth_rates}. We now turn
to characterizing each of them.

\begin{figure}
  \centering
  \subfloat[The Even Process.]{
    \centering
    \begin{tikzpicture}[style=vaucanson, bend angle=15, scale=1, every node/.style={transform shape}]
      \node [state] (A)              {A};
      \node [state] (B) [right of=A] {B};

      \path (A) edge [loop left] node {$\Edge{\half}{0}$} (A)
            (A) edge [bend left] node {$\Edge{\half}{1}$} (B)
            (B) edge [bend left] node {$\Edge{1}{1}$}     (A);
      \path (B) edge [loop right, draw=none] node [opacity=0] {$\Edge{0}{0}$} (B);
    \end{tikzpicture}
    \label{subfig:even}
  } \\
  \subfloat[The Noisy Period Three (NP3) Process.]{
    \centering
    \begin{tikzpicture}[style=vaucanson, bend angle=15, scale=1, every node/.style={transform shape}]
      \node [state] (A) at (0, 0)  {A};
      \node [state] (B) at (240:3) {B};
      \node [state] (C) at (-60:3) {C};

      \path (A) edge [bend right] node [swap] {$\Edge{1}{0}$}     (B)
            (B) edge [bend right] node        {$\Edge{1}{1}$}     (C)
            (C) edge [bend left]  node        {$\Edge{\half}{0}$} (A)
            (C) edge [bend right] node [swap] {$\Edge{\half}{1}$} (A);
    \end{tikzpicture}
    \label{subfig:noisy_period_three}
  } \\
  \subfloat[The Noisy Random Phase-Slip (NRPS) Process.]{
    \centering
    \begin{tikzpicture}[style=vaucanson, bend angle=15, scale=1, every node/.style={transform shape}]
      \foreach \name/\angle/\text in {C/234/C, B/162/B, A/90/A, E/18/E, D/-54/D}
      \node[state,xshift=6cm,yshift=.5cm] (\name) at (\angle:2cm) {$\text$};

      \path (A) edge [loop above] node        {$\Edge{\half}{0}$} (A)
                edge [bend right] node [swap] {$\Edge{\half}{1}$} (B)
            (B) edge [bend right] node [swap] {$\Edge{1}{0}$}     (C)
            (C) edge [bend right] node [swap] {$\Edge{1}{1}$}     (D)
            (D) edge [bend left]  node        {$\Edge{\half}{0}$} (E)
                edge [bend right] node [swap] {$\Edge{\half}{1}$} (E)
            (E) edge [bend right] node [swap] {$\Edge{1}{0}$}     (A);
    \end{tikzpicture}
    \label{subfig:nrps}
  }
  \caption{\EMs\ for the Example Processes.}
  \label{fig:ems}
\end{figure}
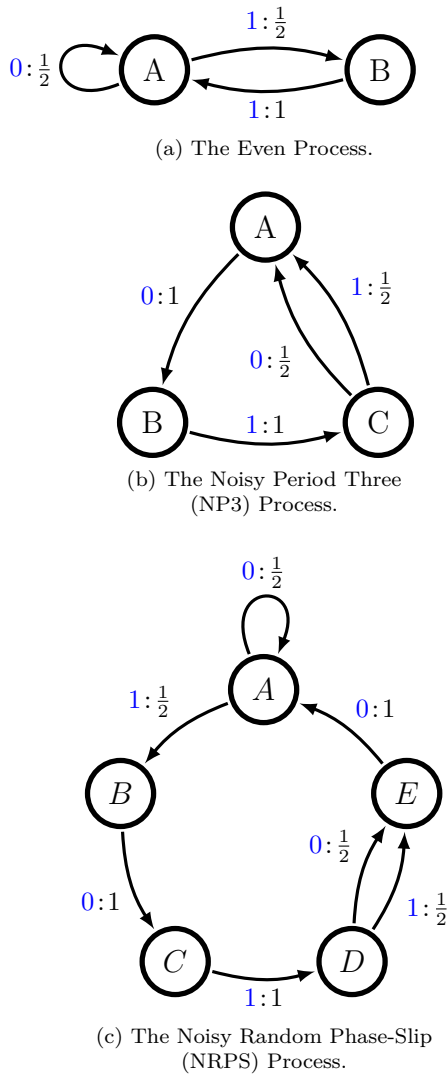

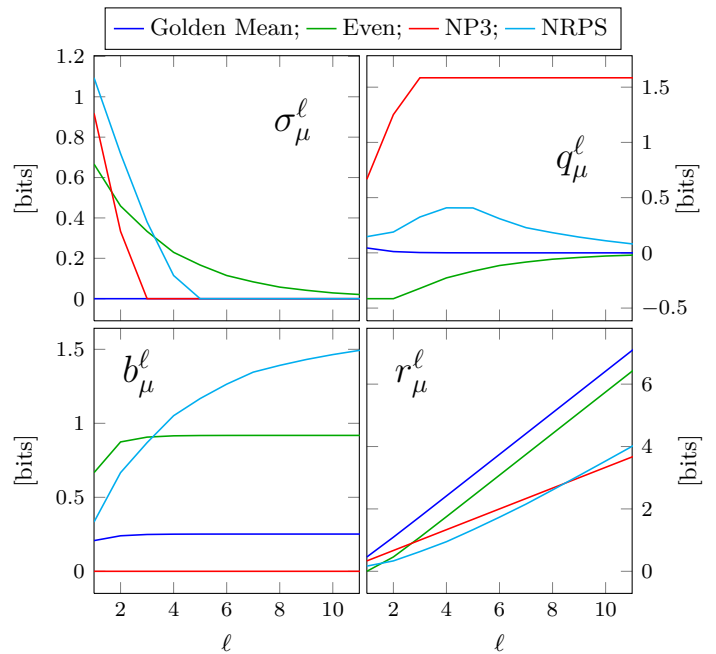
\begin{figure}
  \begin{tikzpicture}
    \pgfplotsset{
      small,
      legend image code/.code={
        \draw[mark repeat=2,mark phase=2] plot coordinates {(0cm, 0cm) (0.3cm, 0cm) (0.4cm, 0cm)};
      },
      every axis plot/.append style={semithick},
      cycle list={blue\\green!66!black\\red\\cyan\\},
    }
    \begin{groupplot}[
        group style={
          group size=2 by 2,
          x descriptions at=edge bottom,
          horizontal sep=0.1cm,
          vertical sep=0.1cm,
        },
        height=0.285\textwidth,
        width=0.285\textwidth,
        xmin=1, xmax=11,
        xlabel={$\ell$},
        ylabel={[bits]},
      ]
      \nextgroupplot [
          legend entries={Golden Mean;, Even;, NP3;, NRPS},
          legend columns=-1,
          legend style={font=\small},
          legend to name=legend]
        \addplot table [x=length, y=s_gm]   {process_data.txt};
        \addplot table [x=length, y=s_even] {process_data.txt};
        \addplot table [x=length, y=s_nemo] {process_data.txt};
        \addplot table [x=length, y=s_nrps] {process_data.txt};
      \nextgroupplot [yticklabel pos=right, ylabel style ={yshift=0.4cm}]
        \addplot table [x=length, y=q_gm]   {process_data.txt};
        \addplot table [x=length, y=q_even] {process_data.txt};
        \addplot table [x=length, y=q_nemo] {process_data.txt};
        \addplot table [x=length, y=q_nrps] {process_data.txt};
      \nextgroupplot
        \addplot table [x=length, y=b_gm]   {process_data.txt};
        \addplot table [x=length, y=b_even] {process_data.txt};
        \addplot table [x=length, y=b_nemo] {process_data.txt};
        \addplot table [x=length, y=b_nrps] {process_data.txt};
      \nextgroupplot [yticklabel pos=right, ylabel style ={yshift=-0.1cm}]
        \addplot table [x=length, y=r_gm]   {process_data.txt};
        \addplot table [x=length, y=r_even] {process_data.txt};
        \addplot table [x=length, y=r_nemo] {process_data.txt};
        \addplot table [x=length, y=r_nrps] {process_data.txt};
    \end{groupplot}
    \node at ($(group c1r1.center)!0.50!(group c1r1.north east)$)                        {{\Large $\sigmamu{\ell}$}};
    \node at ($(group c2r1.center)!0.33!(group c2r1.north east)!0.33!(group c2r1.east)$) {{\Large $\qmu{\ell}$}};
    \node at ($(group c1r2.center)!0.66!(group c1r2.north west)$)                        {{\Large $\bmu{\ell}$}};
    \node at ($(group c2r2.center)!0.66!(group c2r2.north west)$)                        {{\Large $\rmu{\ell}$}};
    \node at ($(group c1r1.north)!0.5!(group c2r1.north)$) [anchor=south, inner sep=2pt] {\ref{legend}};
  \end{tikzpicture}
  \caption{
    Information measures as a function of the present's length $\ell$. Since
	the examples are stationary, finitary processes, both $\sigmamu{\ell}$ and
	$\qmu{\ell}$ converge to zero with increasing $\ell$ and $\bmu{\ell}$
	converges to a constant value of $\EE$. And so, the growth of $\H(\ell)$ is
	entirely captured in $\rmu{\ell}$, and it grows linearly with $\ell$.
  }
  \label{fig:growth_rates}
\end{figure}

We first consider $\sigmamu{\ell}$, seen in \cref{fig:growth_rates}'s upper
left panel. While for each process $\sigmamu{\ell}$ vanishes with increasing
$\ell$, the convergence behaviors differ. The Golden Mean Process is
identically zero at all lengths due to its order-$1$ Markov nature, just noted.
The NRPS Process, with a Markov order of $\MOrder = 5$, has nonzero
$\sigmamu{\ell}$ for $\ell < 5$ and zero $\sigmamu{\ell} = 0$ beyond. Finally,
both the Even and Nemo Processes are infinite-order Markov and so their
$\sigmamu{\ell}$ never exactly vanish, though they converge exponentially fast.
The next section, \cref{subsec:exponential_convergence}, discusses exponential
convergence in more detail.

Next, consider $\qmu{\ell}$ and $\bmu{\ell}$ which, as it turns out, are
closely associated. To see why, first examine the large-$\ell$ limit:
\begin{align*}
\lim_{\ell \to \infty} \I[\LPresent : \LFuture] &= \EE \\
                                                &= \qmu{\infty} + \bmu{\infty}
  ~.
\end{align*}
Second, we can decompose stationary, finite-$\EE$ processes into two classes:
those with state mixing and those without. \emph{State mixing} refers to the
convergence of initial state distributions to a unique invariant distribution,
one that in particular does not oscillate. The GMP, Even, and NRPS Processes
are examples of those with state mixing, while the NP3 Process asymptotic state
distribution is period-$3$, when starting from typical initial distributions.
With state mixing $\Past$ and $\LFuture$ become independent in the
infinite-$\ell$ limit, and so $\qmu{\infty} = 0$ and we conclude $\bmu{\infty}
= \EE$. That is, the entire contribution to excess entropy comes from the bound
information. Without state mixing, though, $\bmu{\ell} = 0$, and so
$\qmu{\infty} = \EE$.

These two classes of convergence behavior are apparent when comparing
\cref{fig:growth_rates}'s upper-right and lower-left panels.  In the
upper-right, $\qmu{\ell}$ converges to zero for the GM, Even, and NRPS
Processes. The NP3 Process, in contrast, limits to a constant value: it's
excess entropy $\EE = \log_2 p$, where $p = 3$ is the period of the internal
state cycle. In the lower-left, $\bmu{\ell}$ limits to constant values for the
three state-mixing processes, while the NP3 Process $\bmu{\ell}$ is zero for
all $\ell$.

Finally, the ephemeral information $\rmu{\ell}$, plotted in \cref{fig:growth_rates}'s
lower-right panel, also depends on whether a process is state mixing or not.
If it is, then:
\begin{align*}
  \H(\ell)   &= \EE + \ell \hmu{} \\
             &= \qmu{\ell} + 2\bmu{\ell} + \rmu{\ell} \\
             &= 0 + 2\EE + \rmu{\ell}
  ~.
\end{align*}
That is, $\rmu{\ell} = -\EE + \ell \hmu{}$.
Though, in the case of no state mixing, we have:
\begin{align*}
  \H(\ell)   &= \qmu{\ell} + 2\bmu{\ell} + \rmu{\ell} \\
             &= \EE + 0 + \rmu{\ell}
  ~.
\end{align*}
That is, $\rmu{\ell} = \ell \hmu{}$.
So, in either case, $\rmu{\ell}$ grows linearly asymptotically with a rate of
$\hmu{}$. If there is state mixing, however, it has a subextensive part equal to
$-\EE$.

\subsection{Exponential Convergence of $\sigmamu{\ell}$}
\label{subsec:exponential_convergence}

One way to classify processes is whether or not an observer can determine the
causal state a process is in from finite or infinite sequence measurements. If
so, then the process is \emph{synchronizable}. All of the previous examples are
synchronizable. References \cite{Trav10a,Trav10b} proved that for any
synchronizable process described by a finite-state HMM, there exist constants
$K > 0$ and $0 < \alpha < 1$ such that:
\begin{align}
  \hmu{}(\ell) - \hmu{} \leq K \alpha^{\ell},
  \quad \textit{for all } \ell \in \mathbb{N}
  ~,
\label{eq:sigma_term_bound}
\end{align}
where:
\begin{align*}
  \hmu{}(\ell) = \H(\ell) - \H(\ell-1)
  ~.
\end{align*}
Note that $\hmu{}(1) = \rhomu{1}$. One well known identity~\cite{Crut01a} is
that the sum of the $\hmu{}(\ell)$ terms is the excess entropy:
\begin{align}
  \EE &= \sum_{k = 1}^{\infty} (\hmu{}(k) - \hmu{}) \\
      &= \rhomu{1} + \sum_{k = 2}^{\infty} (\hmu{}(k) - \hmu{})
  ~.
\end{align}
This provides a new identity \cite{Ara14b}:
\begin{align}
  \sigmamu{1} = \sum_{k=2}^{\infty} (\hmu{}(k) - \hmu{}) ~,
\end{align}
which can be generalized to:
\begin{align}
  \sigmamu{\ell} = \sum_{k=\ell+1}^{\infty} (\hmu{}(k) - \hmu{})
  ~.
\end{align}

\begin{figure}
  \centering
  \includegraphics[scale=0.27]{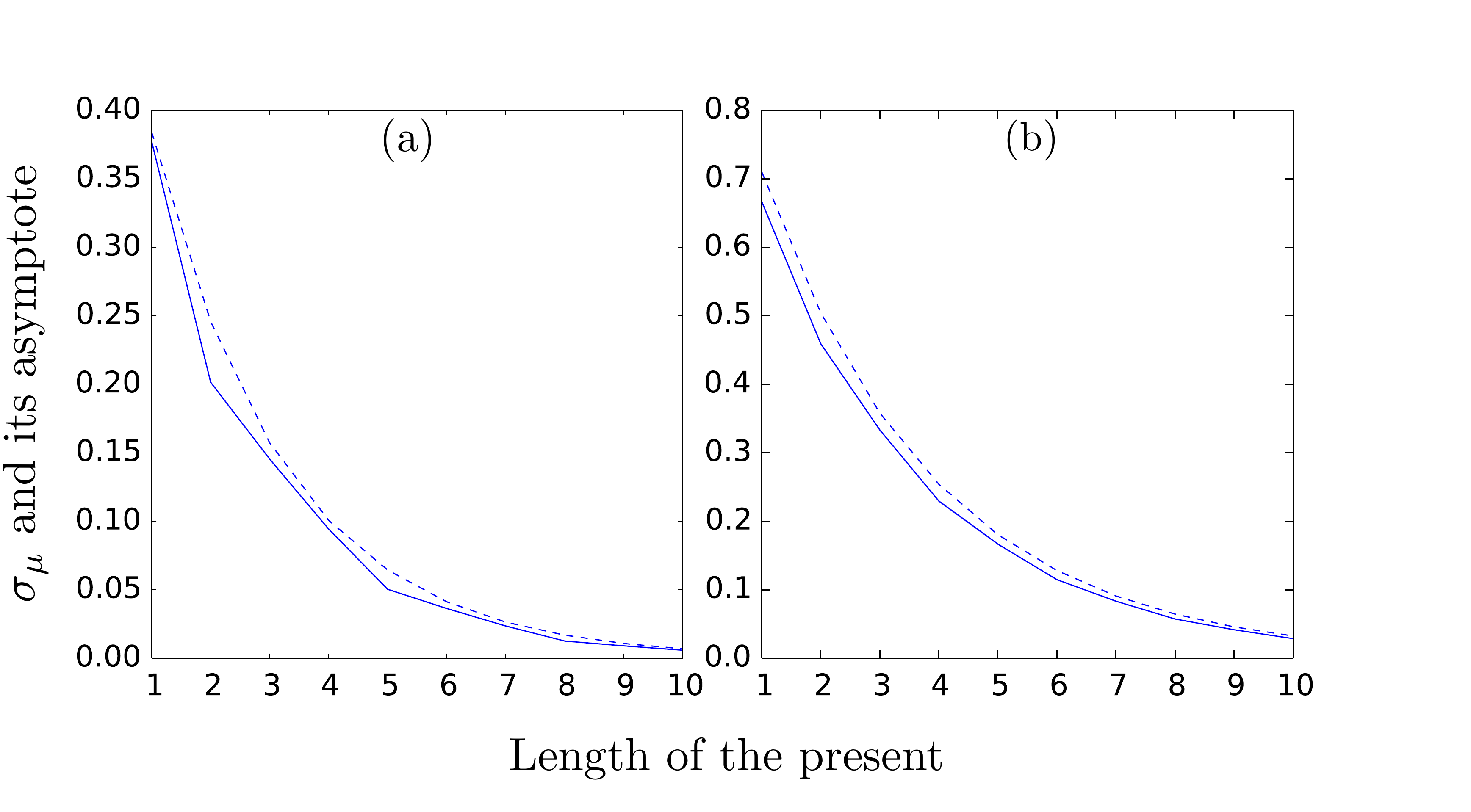}
  \caption{$\sigmamu{\ell}$ (solid) and its asymptote (dashed) for (a)
    Nemo Process with $K = 0.6$ and $\alpha = 0.64$ and (b) Even Process
	with $K = 1$ and $\alpha = 0.71$.
  }
  \label{fig:asym}
\end{figure}

Applying the bound from \cref{eq:sigma_term_bound} to each term, we find:
\begin{align*}
  \sum_{k = \ell+1}^\infty \left( \hmu{}(k) - \hmu{} \right) \leq
    \sum_{k = \ell+1}^\infty  K \alpha^{k}
	~.
\end{align*}
The right-hand side, being a convergent geometric series, yields:
\begin{align*}
  \sigmamu{\ell} \leq  \frac{K \alpha^{\ell+1}}{1 - \alpha}
  ~,
\end{align*}
or simply:
\begin{align}
  \sigmamu{\ell} \leq  K^\prime \alpha^\ell
  ~.
\label{eq:asymp}
\end{align}
We now drop the prime, simplifying the form.  Thus, the elusive information
vanishes exponentially fast for synchronizable processes.

\Cref{fig:asym} compares $\sigmamu{\ell}$ with its best-fit exponential bound
for two different processes: the Nemo Process shown in \cref{fig:nemo} and the
Even Process. For each, the solid line is $\sigmamu{\ell}$ and the dashed is
the fit. Estimated values for the Nemo Process are $K = 0.6$ and $\alpha =
0.64$. The fit parameters for the Even Process are $K = 1.0$ and $\alpha =
0.71$. They were estimated in accordance with the conditions stated for
\cref{eq:sigma_term_bound}. The fits validate the convergence in
\cref{eq:asymp}.

\begin{figure}
  \centering
  \begin{tikzpicture}[style=vaucanson, bend angle=15, scale=1, every node/.style={transform shape}]
    \node [state] (A) at (0, 0)  {A};
    \node [state] (B) at (240:3) {B};
    \node [state] (C) at (-60:3) {C};

    \path (A) edge [loop above] node        {$\Edge{\half}{1}$} (A)
          (A) edge [bend right] node [swap] {$\Edge{\half}{0}$} (B)
          (B) edge [bend right] node        {$\Edge{1}{0}$}     (C)
          (C) edge [bend left]  node        {$\Edge{\half}{0}$} (A)
          (C) edge [bend right] node [swap] {$\Edge{\half}{1}$} (A);
  \end{tikzpicture}
  \caption{The Nemo Process.}
  \label{fig:nemo}
\end{figure}
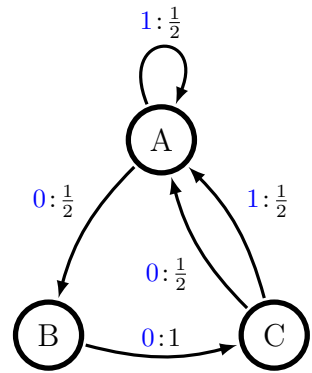

\section{Measures of Emergence?}
\label{sec:related}

The elusive information $\sigmamu{\ell}$ conditions on a present of length
$\ell$. What if we do not condition, simply ignoring the present?  It
becomes the \emph{persistent mutual information} (PMI) \cite{Ball10a,Diak11a}:
\begin{align}
  \PMI{\ell} = \I[\Past : \LFuture]
  ~.
\end{align}
Notably, $\PMI{\infty}$ was offered up as a measure of ``emergence'' in general
complex systems. Our preceding analysis, though, gives a more nuanced view of
this interpretation, especially when emergence is considered in light of
structural criteria introduced previously \cite{Crut93g,Crut92c}. Our framework
reveals that $\PMI{\ell}$ is not an atomic measure; rather it consists of two
familiar components:
\begin{align}
  \PMI{\ell} = \qmu{\ell} + \sigmamu{\ell}
  ~.
\end{align}
Which component is most important? Are both? Which is associated with emergence?
Both?

\Cref{subsec:exponential_convergence} showed that synchronizable processes have
$\sigmamu{\ell} \to 0$. So, for this broad class at least, $\PMI{\infty} =
\qmu{\infty}$. Based on extensive process surveys that we do not report on here,
we conjecture that $\sigmamu{\infty} = 0$ holds even more generally. And so, it
appears that $\PMI{\infty}$ generally is dominated by the multivariate mutual
information $\qmu{\infty}$. Moreover, recalling the analysis of $\qmu{\ell}$ for
the Noisy Period-3 Process shown in the upper-right panel of
\cref{fig:growth_rates}, it appears that $\PMI{\infty}$ is only sensitive to
periodicity or noisy periodicity, giving $\log_2 p$, where $p$ is the period.

As a test of our conjecture that the elusive information vanishes and that
$\PMI{\infty}$ is dominated by $\qmu{\infty}$, we applied our
information-measure estimation methods to
the symbolic dynamics generated by the
Logistic Map of the unit interval as a function of its control parameter $r$.
\Cref{fig:logistic} plots the results.
Indeed, the elusive information does vanish. Thus, we
conclude that $\PMI{\infty}$ is a property of $\qmu{\infty}$.

In addition, our simulation results reproduced those in Ref. \cite{Ball10a}'s
$\PMI{\infty}$ analysis of the Logistic Map; though, their estimation method
for $\PMI{\infty}$ differs markedly. Here, we calculate via the Logistic Map
symbolic dynamics; there, joint distributions over the continuous
unit-interval domain were
used. Both investigations lead to the conclusion that $\PMI{\infty}$ is equal to
(the logarithm of) the number of chaotic ``bands'' cyclically permuted or the
period of the periodic orbit at a given parameter value. In short,
$\PMI{\infty}$ is a measure of \emph{non-mixing} dynamics.

Given the restricted form of structure (periodicity) to which it is sensitive,
$\PMI{\infty}$ cannot be taken as a general measure for detecting the emergence
of organization in complex systems. No matter, though a quarter of a century
old, the statistical complexity \cite{Crut88a}---a direct measure of structural
organization and stored information---continues to fill the role of detecting
emergent organization quite well. Moreover, computational mechanics' \eMs\
directly show what the emergent organization is.

\begin{figure}
  \begin{tikzpicture}
    \pgfplotsset{
      every axis plot/.append style={semithick},
    }
    \begin{axis}[
        legend entries={$\PMI{\infty}$,$\sigmamu{\infty}$,$\qmu{\infty}$},
        legend pos=north west,
    	  x label style={at={(axis description cs:0.5,-0.05)},anchor=north},
    	  y label style={at={(axis description cs:-0.05,.5)},anchor=south},
    	  xlabel={$r$},
    	  ylabel={[bits]},
        xmin=3.2, xmax=4,
        ymin=-0.2, ymax=6,
      ]
      \addplot [blue, very thick] table [x=r, y=pmi] {logistic_pmi_data.dat};
      \addplot [red]              table [x=r, y=smu] {logistic_pmi_data.dat};
      \addplot [green!66!black]   table [x=r, y=qmu] {logistic_pmi_data.dat};
    \end{axis}
  \end{tikzpicture}
\caption{Persistent mutual information
  $\PMI{\infty}$, elusivity $\sigmamu{\infty}$, and multivariate mutual
  information $\qmu{\infty}$ of the Logistic Map symbolic dynamics as a function
  of map control parameter $r$. Recall that the symbolic dynamics does not see
  period-doubling until $r$ is above the appearance of the associated
  superstable periodic orbit. This discrepancy in the appearance of
  periodicity as a function of $r$ does not occur when the map is chaotic.
  (Cf. Fig.~1 of Ref. \cite{Ball10a}.)
  }
\label{fig:logistic}
\end{figure}
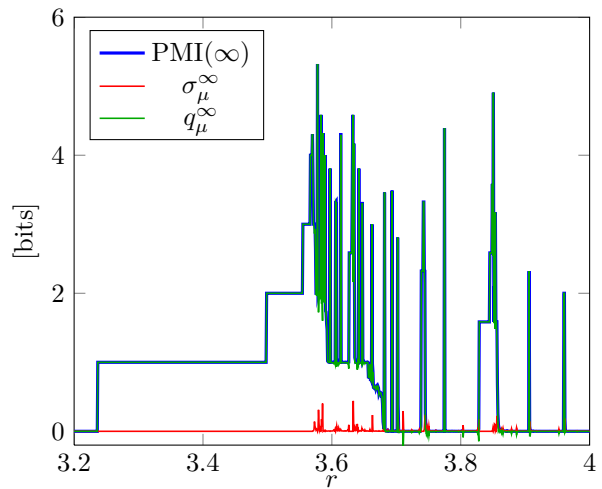


\section{Conclusion}

We first defined the elusive information and developed a closed-form analytic
expression to calculate it from a process's hidden Markov model. The sequel
\cite{Ara14b} shows how to use spectral methods \cite{Crut13a} to develop
alternative closed-form expressions for the elusive information and its
companions.

Investigating how the present shields the past and future is essentially a
study of what Markov order means for structured processes. It gives some
insight into the process of modeling building and even general concerns about
emergence in complex systems. First, it is a common ground on which to contrast
structural inference and emergence, showing that we should not conflate these
two distinct questions. Perhaps most constructively, though, it sheds light on
the challenges of inference for complex systems. In particular, when
$\sigmamu{\ell} > 0$ sequence statistics are inadequate for modeling and so we
must use state-based models to properly, finitely represent a process's
internal organization.

We showed how present observables typically do not contain all of the information that correlates the past and the future. One consequence is that instantaneous measurements are not enough. This means, exactly, that Markov chain models of complex physical systems are fundamentally inadequate, though eminently helpful and simplifying when they are appropriate representations.  The larger consequence is that we must build state-based models and not use mere look-up tables or sequence histograms. And this means, in turn, that monitoring only prediction performance is inadequate. We must also monitor model complexity, not as an antidote to over-fitting, but as a fundamental goal for both prediction and understanding hidden mechanisms.

\acknowledgments

We thank the Santa Fe Institute for
its hospitality during visits. JPC is an SFI External Faculty member. This
material is based upon work supported by, or in part by, the U. S. Army
Research Laboratory and the U. S. Army Research Office under contracts
W911NF-13-1-0390 and W911NF-13-1-0340.


\bibliographystyle{unsrt}
\bibliography{chaos}

\end{document}

%% file: vaucanson.tikz
\tikzstyle{vaucanson}=[
  node distance=3cm,
  bend angle=15,
  auto,
  every loop/.style={},
  every edge/.style={->,draw=black,line width=1.2,>=latex,shorten <=1pt, shorten >=1pt},
  every state/.style={draw=black,line width=2,font=\large},
  loop right/.style={right,out=22,in=-22,loop},
  loop above/.style={above,out=112,in=68,loop},
  loop left/.style={left,out=202,in=158,loop},
  loop below/.style={below,out=292,in=248,loop},
  loop above right/.style={right,out=67,in=23,loop},
  loop above left/.style={left,out=157,in=113,loop},
  loop below left/.style={left,out=247,in=203,loop},
  loop below right/.style={right,out=337,in=293,loop},
  binode/.style={minimum size=1cm,inner sep=0pt},
]

\tikzset{
  mystate/.style={circle,draw,fill=black}
}

\makeatletter
\newcommand{\binode}[1][state]{%
  \@ifnextchar[{\binode@i[{#1}]}{\binode@i[{#1}][{yellow},{gray!70}]}%
}
\def\binode@i[#1][#2,#3]{%
  \@ifnextchar({\binode@ii[{#1}][{#2},{#3}]}{\binode@ii[{#1}][{#2},{#3}]({})}%
}
\def\binode@ii[#1][#2,#3](#4){%
  \@ifnextchar[{\binode@iii[{#1}][{#2},{#3}]({#4})}{\binode@iii[{#1}][{#2},{#3}]({#4})[{}]}%
}
\def\binode@iii[#1][#2,#3](#4)[#5]#6{%

  \node[#1,binode] (#4) [#5] {#6};
  \node[yshift=-10pt] (#4SS) at (#4.270) {};
  \node[xshift=-10pt] (#4WW) at (#4.180) {};
  \node[xshift=-10pt] (#4NN) at (#4.180) {};
  \node[xshift=-10pt,yshift=10pt] (#4NW) at (#4.135) {};
  \begin{scope}
    \path[clip] (#4.255) -- +(-.8cm,0cm) -- +(-.8cm,1cm) -- (#4.75) -- cycle;
    \node[#1,fill=#2,binode] (#4f) [#5] {#6};
  \end{scope}
  \begin{scope}
    \path[clip] (#4.75) -- +(.8cm,0cm) -- +(.8cm,-1cm) -- (#4.255) -- cycle;
    \node[#1,fill=#3,binode] (#4r) [#5] {#6};
  \end{scope}
  \node {} 
}
\makeatother

\providecommand{\Symbol}[1]{\textcolor{blue}{#1}}
\providecommand{\Edge}[2]{\Symbol{#2}\!:\!#1}

\definecolor{FCSA}{RGB}{141,211,199}
\definecolor{FCSB}{RGB}{255,255,179}
\definecolor{RCSC}{RGB}{185,138,196}
\definecolor{RCSD}{RGB}{231,143,111}
\definecolor{RCSE}{RGB}{128,177,211}

%% file: cmechabbrev.tex
\newcommand{\eM}     {\mbox{$\epsilon$-machine}}
\newcommand{\eMs}    {\mbox{$\epsilon$-machines}}
\newcommand{\EM}     {\mbox{$\epsilon$-Machine}}
\newcommand{\EMs}    {\mbox{$\epsilon$-Machines}}

\newcommand{\Process}{\mathcal{P}}

\newcommand{\MeasAlphabet}      {\mathcal{A}}
\newcommand{\MeasSymbol}   { {X} }
\newcommand{\meassymbol}   { {x} }

\newcommand{\Past}      { \smash{\overleftarrow {\MeasSymbol}} }
\newcommand{\past}      { \smash{\overleftarrow {\meassymbol}} }

\newcommand{\Future}    { \smash{\overrightarrow{\MeasSymbol}} }

\newcommand{\CausalState}       { \mathcal{S} }

\newcommand{\causalstate}       { \sigma }
\newcommand{\CausalStateSet}    { \boldsymbol{\CausalState} }
\newcommand{\AlternateState}    { \mathcal{R} }

\newcommand{\alternatestate}    { \rho }

\newcommand{\CausalEquivalence} { {\sim}_{\epsilon} }

\newcommand{\Prob}      {\Pr} 

\newcommand{\hmu}               {h_\mu}
\newcommand{\EE}                {{\bf E}}

\newcommand{\ProcessAlphabet}   {\MeasAlphabet}

\newcommand{\forward}{+}
\newcommand{\reverse}{-}
\newcommand{\forwardreverse}{\pm} 

\newcommand{\FutureCausalState} { {\CausalState}^{\forward} }

\newcommand{\PastCausalState}   { {\CausalState}^{\reverse} }

\newcommand{\lastindex}[2]{
  \edef\tempa{0}
  \edef\tempb{#2}
  \ifx\tempa\tempb
    \edef\tempc{#1}
  \else
    \edef\tempa{0}
    \edef\tempb{#1}
    \ifx\tempa\tempb
      \edef\tempc{#2}
    \else
      \edef\tempc{#1+#2}
    \fi
  \fi
  \tempc
}

\newcommand{\MOrder}{R}

\newcommand{\rhomu}{\rho_\mu}
\newcommand{\rmu}{r_\mu}
\newcommand{\bmu}{b_\mu}
\newcommand{\qmu}{q_\mu}

\newcommand{\sigmamu}{\sigma_\mu}

\newcommand{\I}{\mathbf{I}}

\newcommand{\CSjoint}[1][,]{
   \edef\tempa{:}
   \edef\tempb{#1}
   \ifx\tempa\tempb
      \ensuremath{\FutureCausalState\!#1\PastCausalState}
   \else
      \ensuremath{\FutureCausalState#1\PastCausalState}
   \fi
}
\newcommand{\CSjointKL}[3][,]{
   \edef\tempa{:}
   \edef\tempb{#1}
   \ifx\tempa\tempb
      \ensuremath{\FutureCausalState_{#2}\!#1\PastCausalState_{#3}}
   \else
      \ensuremath{\FutureCausalState_{#2}#1\PastCausalState_{#3}}
   \fi
}

\newif\ifpm
\edef\tempa{\forwardreverse}
\edef\tempb{\pm}
\ifx\tempa\tempb
   \pmfalse
\else
   \pmtrue
\fi

\newcommand{\MeasSymbols}[2] {\MeasSymbol_{#1:#2}}

\renewcommand{\Past}{\MeasSymbols{}{0}}
\newcommand{\Present}{\MeasSymbol_0}
\renewcommand{\Future}{\MeasSymbols{1}{}}

\ifcsname mathclap\endcsname
  \relax
\else
  \def\clap#1{\hbox to 0pt{\hss#1\hss}}

  \def\mathclap{\mathpalette\mathclapinternal}

  \def\mathclapinternal#1#2{%
  \clap{$\mathsurround=0pt#1{#2}$}}
\fi